\documentclass[pra,aps,twocolumn,preprintnumbers,showpacs,superscriptaddress]{revtex4}
\usepackage{latexsym,amssymb,amsfonts,amsmath,graphicx,bbm}
\usepackage[usenames,dvipsnames]{color}
\newcommand{\eh}{\mathrm{e}}

\newcommand{\Tr}{\mathrm{Tr}}
\newcommand{\dd}{\; \mathrm{d}}

\newcommand{\ket}[1]{ | \, #1 \rangle}

\newcommand{\bra}[1]{ \langle #1 \,  |}

\newcommand{\od}[2]{\frac{\mathrm{d} #1}{\mathrm{d} #2}}

\newcommand{\eqr}[1]{Eq.~(\ref{#1})}
\newcommand{\fir}[1]{Fig.~\ref{#1}}
\newcommand{\secr}[1]{Sec.~\ref{#1}}
\newcommand{\apr}[1]{Appendix~\ref{#1}}

\newcommand{\an}[1]{\hat{#1}}
\newcommand{\cre}[1]{\hat{#1}^\dag}

\begin{document}

\title{Impurity transport through a strongly interacting bosonic quantum gas}
\author{T. H. Johnson}
\email{t.johnson1@physics.ox.ac.uk}
\affiliation{Clarendon Laboratory, University of Oxford, Parks Road, Oxford OX1 3PU, United Kingdom}
\author{S. R. Clark}
\affiliation{Centre for Quantum Technologies, National University of Singapore, 3 Science Drive 2, Singapore 117543, Singapore}
\affiliation{Clarendon Laboratory, University of Oxford, Parks Road, Oxford OX1 3PU, United Kingdom}
\affiliation{Keble College, University of Oxford, Parks Road, Oxford OX1 3PG, United Kingdom}
\author{M. Bruderer}
\affiliation{Fachbereich Physik, Universit\"{a}t Konstanz, D-78457 Konstanz, Germany}
\author{D. Jaksch}
\affiliation{Clarendon Laboratory, University of Oxford, Parks Road, Oxford OX1 3PU, United Kingdom}
\affiliation{Centre for Quantum Technologies, National University of Singapore, 3 Science Drive 2, Singapore 117543, Singapore}
\affiliation{Keble College, University of Oxford, Parks Road, Oxford OX1 3PG, United Kingdom}

\date{\today}
%%%%%%%%%%%%%%%%%%%%%%

\begin{abstract}
Using near-exact numerical simulations we study the propagation of an
impurity through a one-dimensional Bose lattice gas for varying bosonic
interaction strengths and filling factors at zero temperature. The
impurity is coupled to the Bose gas and confined to a separate tilted lattice. The precise nature of the transport of the impurity is specific to the excitation
spectrum of the Bose gas which allows one to measure properties of the Bose gas non-destructively, in principle, by observing the impurity; here we focus on the spatial and momentum distributions of the impurity as well as its reduced density matrix. For instance we show it is possible to determine whether the Bose gas is commensurately filled as well as the bandwidth and gap in its excitation spectrum. Moreover, we show that the impurity acts as a witness to
the cross-over of its environment from the weakly to the strongly
interacting regime, i.e., from a superfluid to a Mott insulator or
Tonks-Girardeau lattice gas and the effects on the impurity in both of these strongly-interacting regimes are clearly distinguishable. Finally, we find that the spatial coherence of the
impurity is related to its propagation through the Bose gas, giving an
experimentally controllable example of noise-enhanced quantum transport.
\end{abstract}
%%%%%%%%%%%%%%%%%%%%%%

\pacs{67.85.-d, 05.60.Gg, 47.60.-i, 72.10.-d}

\maketitle
%%%%%%%%%%%%%%%%%%%%%%

\section{Introduction}
The transport of an impurity through a Bose gas has received much attention
due to both its inherent appeal as an example of nonequilibrium quantum
dynamics and its importance in mimicking the transport of an electron in a
conductor~\cite{Lewenstein2007}. The advantage of a cold atom and optical
lattice setup is its ability to realize idealized Hamiltonians and fine-tune
interactions over large parameter regimes, often beyond those found in
condensed matter systems~\cite{Bloch2008}. In this vein, we
extend the large body of theoretical work on the transport of an impurity
(partly subject to static forcing) through a
superfluid~\cite{Astrakharchik2004,Ponomarev2006,Klein2007,Bruderer2008,Tempere2009,Bruderer2010,Privitera2010}
by considering strongly-interacting bosons in one dimension up to the Mott insulator and
Tonks-Girardeau limits, whereupon the bosons exhibit
fermionic characteristics~\cite{Cazalilla2003}. Recently, impurity transport
in this context has received interest due to its successful realization by two
experiments, one using a species specific dipole potential \cite{Catani2011} and another using gravity to provide a static force for the impurities~\cite{Palzer2009,Rutherford2010}. In this article we look at how an impurity subjected to a one-dimensional
tilted optical lattice potential moves through a bosonic lattice gas in
diverse regimes. We simulate the system non-perturbatively and near-exactly
at zero temperature using the matrix product based time-evolution block decimation (TEBD)
algorithm~\cite{Vidal2003,Vidal2004}.
%%%%%%%%%%%%%%%%%%%%%%
\begin{figure*}[tbh]
\includegraphics[width=17.8cm]{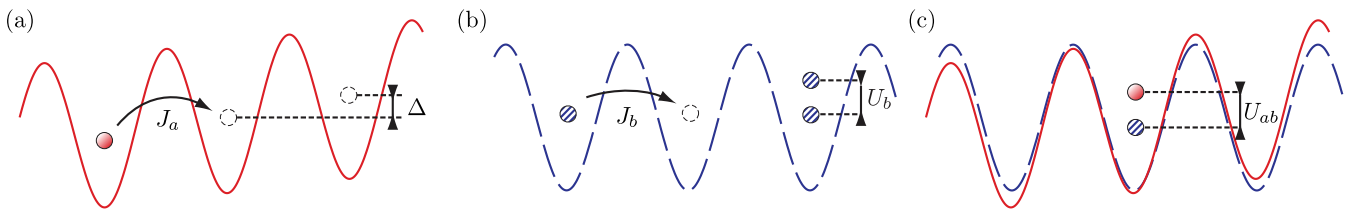}
\caption{\label{fig:model}(Color online) Schematic diagram for the (a) impurity, (b) Bose gas and (c) interaction Hamiltonians.}
\end{figure*}
%%%%%%%%%%%%%%%%%%%%%%

This work also relates to progress made in probing cold atoms
in optical lattices. For example, the excitation properties of a Bose gas in
an optical lattice can be analyzed using Bragg spectroscopy via two photon
scattering~\cite{Stenger1999} and lattice depth
modulation~\cite{Stoferle2004,Haller2010} --- the experiment
in~\cite{Stoferle2004} having been later successfully simulated
in~\cite{Kollath2006,Clark2006}. Studying the transport of a Bose gas after
imparting a momentum on it using a magnetic field~\cite{Haller2010} or
tilting the lattice~\cite{Greiner2002} has allowed experimentalists to
successfully reveal signatures of phase transitions or determine excitation
spectra, respectively. Recently an impurity has been used in experiments as
a coherent probe for large quantum systems in order to extract
impurity-boson collision parameters~\cite{Weber2010} and interaction
strengths~\cite{Will2011}.

We here combine and extend these two ideas of probing a system using impurities and transport; we discuss
how the motion of an impurity depends on its environment and then using our simulations we show
what information can be revealed about the environment non-destructively by observing the impurity. As our example, we consider the environment to be a Bose lattice gas and demonstrate that by analyzing the transport of the impurity one can determine the bandwidth and
the gap in the excitation spectrum of the Bose gas. In the case of a commensurately filled Bose gas the impurity propagation provides a clear signature of the superfluid to Mott insulator transition in the form of a sharp quench of the expected current. The strong dependence on the filling allows one to determine non-destructively whether the bosons have been prepared in a commensurately filled state, as desired in some applications of quantum information processing.

The structure of this article is as follows: In \secr{sec:setup} we
develop the theoretical framework underlying the dynamics of impurities in a tilted lattice, first decoupled
and then coupled via interactions to an environment. Subsequently, we
specialize to the case of an environment constituting a Bose lattice gas.
In \secr{sec:current} we simulate the propagation of the impurity. Specifically, in Secs. \ref{sec:esakitsu} and \ref{sec:sftomi} we analyze the displacement of the impurity in an incommensurately and commensurately filled Bose gas, respectively, and show
that deviations from the basic Esaki-Tsu dependence on tilt clearly reveal information about the gas. In \secr{sec:momdist} we show how the interaction of the
impurity with the Bose gas affects the momentum distribution of the impurity. Then in \secr{sec:denmat} we
discuss the decoherence of the impurity as an example of noise-enhanced
transport. We
make our conclusions in \secr{sec:conclusion} and leave a brief review of
the TEBD algorithm to \apr{sec:TEBD}.
%%%%%%%%%%%%%%%%%%%%%%

\section{Theoretical framework}
\label{sec:setup}
%%%%%%%%%%%%%%%%%%%%%%

\subsection{Impurities in a tilted lattice}
\label{sec:impurity} In our system we have one or a few impurities confined
to a tilted optical lattice which allows them to move along one spatial
dimension only, depicted in \fir{fig:model}(a). Such an optical lattice can be
created for the impurities by subjecting them to both an optical lattice and
a static forcing, e.g., gravity~\cite{Palzer2009}, or by chirping the
frequency difference between the counter-propagating lasers making up the
standing wave of an optical lattice along the direction of
motion~\cite{Madison1997}. For sufficiently deep lattices and low
temperatures solely the lowest Bloch band is occupied and only tunneling
between nearest neighbor sites must be considered. The Hamiltonian of the
impurities is of the form
\begin{equation}\label{eq:ha}
    \hat{H}_a = -J_a \sum_{\langle i,j\rangle} \hat{a}^\dagger_i \hat{a}_j + \Delta \sum_i i\,\hat{a}^\dagger_i \hat{a}_i \, ,
\end{equation}
where $\langle i,j \rangle$ denotes nearest neighbor sites $i$ and $j$,
$\hat{a}^\dagger_i$ ($\hat{a}_i$) is the creation (annihilation) operator
for an impurity in a Wannier state~\cite{Wannier1937} localized at site $i$,
separated in energy and distance from its neighbors by the Bloch energy $\Delta$ and the
lattice constant $a$. An additional condition for the above
Hamiltonian to hold is that the separation of the Bloch bands must be much
greater than the Bloch energy. Note that the impurities could be either
bosonic or fermionic; we assume them to be sufficiently dilute so that
impurity-impurity interactions and their quantum statistics have negligible
effects.

Without a further interaction with an environment it is not possible for an impurity to
dissipate energy and therefore alter its expected position. The eigenstates of
the Hamiltonian in \eqr{eq:ha} are centered at each of the lattice
sites $i$ and have energies $\Delta i$ forming the so-called
Wannier-Stark ladder of states. Introducing the width $\Lambda = 2J_a/\Delta$ the creation operator for a Wannier-Stark state is defined by $\cre{d}_i = \sum_j \mathcal{J}_{j-i} (\Lambda)
\cre{a}_j$ where $\mathcal{J}_n (\Lambda)$ is a Bessel function of the first kind of
order $n$. As the states are separated by an energy $\Delta$ a superposition
of them oscillates at a frequency $\Delta / \hbar$.

This Wannier-Stark picture is consistent with the solution of the
semi-classical equations for an impurity in a lattice described by a
wave-packet of well defined quasi-momentum (therefore spatially it must be
spread over many lattice sites)~\cite{Ashcroft1976}. In such a framework the
effect of the tilt is a constant drift of the quasi-momentum of each of the
impurities at rate $\dot{k} = \Delta / \hbar a$.
Together with the semi-classical equation of motion for the group velocity of a wave-packet
\begin{equation}\label{eq:group velocity}
v(k) = \frac{1}{\hbar} \frac{\partial E}{\partial k},
\end{equation}
and the single particle energy spectrum in a lattice $E(k) = 2J_a( 1-  \mathrm{cos}ka)$ this results in a group velocity that is sinusoidal in time. Thus an impurity undergoes harmonic motion with
frequency $\Delta / \hbar$ but there is no net drift down the lattice. Although originally derived
semi-classically, these Bloch oscillations are a purely coherent phenomenon, can be arrived at
quantum-mechanically and are rigorously connected to the Wannier-Stark picture~\cite{Rossi1998}. In \fir{fig:blochesakitsu}(a) we show the coherent Bloch oscillation of an impurity initially located in a Wannier state.

We note that Bloch oscillations in semi-conductors are usually
obscured by large scattering rates, though they have been observed in
superlattices where the Bloch frequencies $\Delta / \hbar$ achievable are much higher due
to larger lattice spacings \cite{Leo2003}. Bloch oscillations are also
observable in analogous optical lattice setups, as we find here, due to
the much smaller scattering rates.
%%%%%%%%%%%%%%%%%%%%%%

\subsection{Impurities coupled to an environment}
\label{sec:impuritycurrent} If the impurities are coupled to an
environment then this introduces a mechanism through which an impurity can
dissipate energy and fall down the lattice. Esaki and Tsu~\cite{Esaki1970},
using the semi-classical Bloch oscillation picture, derived an expression
for the average current in the relaxation-time approximation. The main
assumption is that an impurity suffers collisions at some rate $1/ \tau$
that resets its momentum distribution to that of thermal equilibrium (we
assume zero-temperature here). The resulting expression for the average
velocity of each of the impurities is then
\begin{equation}\label{eq:esaki-tsu}
\langle v \rangle = \frac{2aJ_a}{\hbar} \frac{\tau \Delta / \hbar}{(\tau \Delta / \hbar)^2 + 1} ,
\end{equation}
describing a linear (ohmic) dependence on forcing at small $\tau \Delta$
followed by negative differential conductance as $(\tau \Delta)^{-1}$ when
$\tau \Delta \gg 1$. This dependence is plotted in \fir{fig:blochesakitsu}(b). In this semi-classical framework the negative differential current
arises if the timescale of the collisions is considerably larger than
the period of Bloch oscillations, in which case the velocity of the impurity
is not always in the same direction and averages out between collisions. Negative differential
conductance, as with Bloch oscillations, is usually obscured in semiconductors but
due to the much larger values of $\tau \Delta$ achievable it has been observed in optical lattices \cite{Roati2004} and
superlattices \cite{Sibille1990}.
%%%%%%%%%%%%%%%%%%%%%%
\begin{figure}[tb]
\includegraphics[width=8 cm]{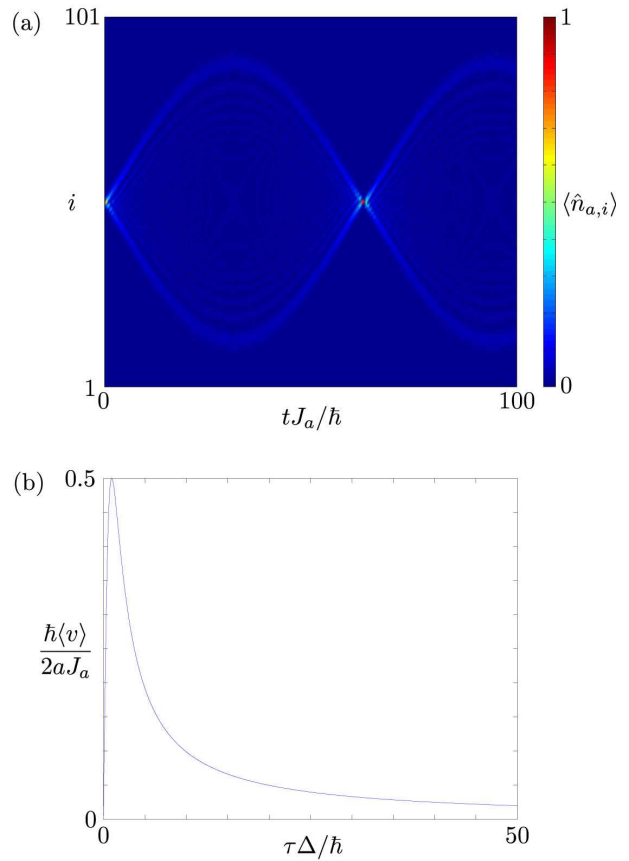}
\caption{\label{fig:blochesakitsu}(Color online) {\em The motion of a single impurity in a tilted lattice with and without a scattering process}. (a) The impurity decoupled from any environment will coherently Bloch oscillate at frequency $\Delta / \hbar$ but its expected position will not drift~\cite{Thomen2004,Klein2007}. Here we plot the evolution in time $t$ for just over one and a half Bloch oscillations. The impurity is initially localized in a Wannier state at the middle site of the $101$-site lattice. We set the tilt to be $\Delta / J_a = 0.1$ and $\langle \hat{n}_{a,i} \rangle$ is the impurity density at lattice site $i$. (b) Introducing a scattering mechanism under the relaxation-time approximation results in the Esaki-Tsu dependence of drift on tilt in \eqr{eq:esaki-tsu}. We plot this dependence here.}
\end{figure}
%%%%%%%%%%%%%%%%%%%%%%

Clearly the relaxation-time approximation is not always valid and the actual
current will depend on the details of the environment and its interaction
with the impurity. However, it has been found that at least qualitatively an
Esaki-Tsu-like dependence of current on forcing is obeyed in many cases: for
density-density interactions between an impurity and a Bose lattice gas at
infinite temperature~\cite{Kolovsky2008}; when an impurity scatters from a
set of different fixed impurities with scattering
resulting in a Fermi distribution in momentum space~\cite{Wacker1998}; and
where an impurity scatters on phonons~\cite{Bryxin1972}.

Turning this on its head, the precise way in which the current of the
impurity deviates from the Esaki-Tsu dependence reveals information about
the environment and impurity-environment interaction. For example, the ability to excite optical phonons in the environment at a single frequency result in resonances in the current of an impurity at certain
tilts dependent on this frequency~\cite{Bryxin1972}. The interaction of impurities
with a superfluid in a lattice, which has a divergence in the density of
phonon states at the band edge, also causes peaks at forcings related
to fractions of the bandwidth~\cite{Bruderer2010}.

We now consider in which way properties of the environment manifest
themselves in the current of the impurities, approaching the problem using
both the semi-classical and Wannier-Stark pictures.
%%%%%%%%%%%%%%%%%%%%%%

\subsubsection{Scattering in momentum space}
\label{sec:scattering} First, we use the semi-classical approach where the
distribution of impurities in quasi-momentum space $g(k)$ obeys the
Boltzmann equation~\cite{Ashcroft1976}
\begin{equation}\label{eq:boltzmanneq}
\od{g(k)}{k} = \frac{\hbar a}{\Delta} \int \dd q \left[ W(q,k) g(q) - W(k,q) g(k) \right],
\end{equation}
which is valid for homogeneous systems of impurities in the steady
state and where $W(q,k)$ is the rate of incoherent scattering between
quasi-momenta $q$ and $k$. The average velocity of each impurity is then
the expected value of the group velocity of wave-packets distributed
according to $g(k)$,
\begin{equation}\label{eq:semiclassicalcurrent}
\langle v \rangle = \int \dd k g(k) v(k) ,
\end{equation}
where the group velocity for a particular quasi-momentum is defined in
\eqr{eq:group velocity}. For small enough tilts the quasi-momentum will be a
very slowly changing quantity and hence for
weak interactions with the environment we may approximate the
incoherent scattering rates using Fermi's golden rule as
\begin{equation}\label{eq:fgr1} \nonumber
\begin{aligned}
W(q,k) = &\frac{2 \pi}{\hbar} \sum_{n_b} \left | \bra{k} \bra{n_b} \hat{H}_{ab} \ket{q} \ket{0_b} \right |^2  \\ 
 &\times \delta \left( E_{n_b} - E_{0_b} - 2J_a [ \mathrm{cos}qa - \mathrm{cos}ka ] \right) .
\end{aligned}
\end{equation}
Here we have assumed the environment to be sufficiently large and
cold so that before scattering we may take it to be in the
ground state $\ket{0_b}$ of energy $E_{0_b}$. The result of the
interaction of the impurity with the environment, for which the Hamiltonian
is given by $\hat{H}_{ab}$, is to scatter the environment to some excited
state $\ket{n_b}$ with energy $E_{n_b}$.

Particularly important for our system is the case of impurities coupled
to a many-body environment, here the Bose gas, via a density-density
interaction. For low energies, the interaction takes the explicit
form
\begin{equation}\label{eq:hab}
\hat{H}_{ab} = U_{ab} \sum_i \hat{a}^\dagger_i \hat{a}_i \hat{b}^\dagger_i \hat{b}_i ,
\end{equation}
where $U_{ab}$ is its characteristic energy and $\hat{b}^\dagger_i$
($\hat{b}_i$) is the creation (annihilation) operator for a particle in the
environment at lattice site $i$. The corresponding quasi-momentum
scattering rates are given by~\cite{Timmermans1998}
\begin{equation}\label{eq:fgr4}
W(q,k) = \frac{2 \pi U_{ab}^2}{\hbar L} S(q-k, - 2J_a [ \mathrm{cos}qa - \mathrm{cos}ka ] )  ,
\end{equation}
where $L$ is the length of the system, $S(q, \omega)$ is the zero-temperature dynamic structure factor of the environment
\begin{align}
\label{eq:dsf} \nonumber
S(q, \omega) &= \sum_{n_b} | \bra{n_b} \hat{\rho}_q \ket{ 0_b} |^2 \delta ( E_{n_b} - E_{0_b}  - \hbar \omega ) ,
\end{align}
and $\hat{\rho}_q$ is the Fourier transform of the particle
density of the environment. Hence the impurity current and, more directly, the momentum
distribution at small tilts are governed by the dynamic structure factor
of the environment.
%%%%%%%%%%%%%%%%%%%%%%

\subsubsection{Transitions between Wannier-Stark states}
\label{sec:wannierstark} At larger lattice tilts it is convenient to
work in the basis of Wannier-Stark states and to consider the incoherent
transitions between these. The dynamics
of the impurities is described by the master equation
\begin{equation} \nonumber
  \od{P_i (t)}{t} = \sum_{j \ne i}\left[ W_{ji} P_j (t) - W_{ij} P_i (t)  \right] ,
\end{equation}
with the incoherent hopping rates $W_{ij}$. The probability $P_i$ that the impurity occupies
Wannier-Stark states at site $i$ is normalized to $\sum_i P_i(t) = 1$. The corresponding average velocity then
reads
\begin{equation}\label{eq:fgr3} \nonumber
    \langle v \rangle = \sum_i a i \od{P_i (t)}{t}.
\end{equation}
Note that in an infinite homogeneous system the average velocity would be
$\langle v \rangle = \sum_j a (j-i) W_{ij}$. Again, for weak coupling to the
environment the transition rates $W_{ij}$ between different Wannier-Stark
states are given by Fermi's golden rule as
\begin{equation} \label{eq:fgr2}
\begin{aligned}
W_{ij} = &\frac{2 \pi}{\hbar} \sum_{n_b} \left | \bra{j} \bra{n_b} \hat{H}_{ab} \ket{i} \ket{0_b} \right |^2 \\
&\times \delta \left( E_{n_b} - E_{0_b} + [i-j] \Delta \right) .
\end{aligned}
\end{equation}
In this picture negative differential current arises due to the increasing
localization of the Wannier-Stark states: as $\Delta$ increases, the width
of the Wannier-Stark states decreases as $1/\Delta$ resulting in a
suppression of their overlaps and thus the motion of the impurity.

Inserting the interaction Hamiltonian in \eqr{eq:hab} into \eqr{eq:fgr2} we arrive at the environment correlation function
\begin{equation}\label{eq:besselcorr}
\begin{aligned}
W_{ij} = &\frac{2 \pi U_{ab}^2}{\hbar}
\sum_{n_b} \Big| \bra{n_b} \sum_\ell \mathcal{J}_{\ell-j}(\Lambda)
\mathcal{J}_{\ell-i}(\Lambda)\hat{b}^\dagger_\ell \hat{b}_\ell \ket{0_b} \Big|^2 \\
&\times \delta \left( E_{n_b} - E_{0_b} + [i-j] \Delta \right) ,
\end{aligned}
\end{equation}
which determines the current at larger tilts. From this we immediately see two effects that we expect to be visible in the impurity
current. First, consider the case where the environment has a gap $G$
in its excitation spectrum such that $E_{n_b} - E_{0_b} \ge G$.
This means that the delta function in \eqr{eq:besselcorr} will be zero
unless $\Delta \ge G/(j-i)$ and hence the contributions to the current due
to hopping between Wannier-Stark states shifted by $\ell$ sites will only be
made at tilts above $\Delta = G/\ell$. Second, if the excitation
spectrum of the environment has a finite bandwidth such that there
are no states above an energy $B$, i.e., $E_{n_b} - E_{0_b} \le B$,
then we expect $\ell$-site contributions to the current to be suppressed at
tilts above $\Delta = B/\ell$. Hence the deviation of the dissipation process from the
relaxation-time approximation results in the dependence of the
impurity current on $\Delta$ exhibiting sharp features that clearly reveal information about the environment; we can infer these environmental properties by observing the impurities.
%%%%%%%%%%%%%%%%%%%%%%

\subsection{Properties of the Bose gas environment}
Our specific environment consists of a Bose gas trapped in a horizontal (non-tilted)
optical lattice in which the bosons are confined to move in the same single dimension as the impurities. As
for the impurities we consider temperatures to be low enough such that only
the lowest Bloch band need be considered. In such a case the bosons are
described by the Bose-Hubbard model~\cite{Jaksch1998}
\begin{equation}\label{eq:hb}
\hat{H}_b = - J_b \sum_{\langle i,j \rangle} \hat{b}^\dagger_i \hat{b}_j + \frac{U_b}{2} \sum_i \hat{b}^\dagger_i \hat{b}^\dagger_i \hat{b}_i \nonumber
\hat{b}_i
\, ,
\end{equation}
where the operator $\hat{b}^\dagger_i$ ($\hat{b}_i$) creates (annihilates) a
boson in a Wannier state localized at site $i$ and the lattice parameter
$a$ is the same as that for the impurities. $J_b$ and $U_b$ are parameters
that determine the hopping between neighboring sites and the on-site
interaction, respectively, and can be tuned experimentally by adjusting the
laser parameters or using Feshbach resonances~\cite{Jaksch1998,Bloch2008}.

Let us briefly review some properties of the Bose gas which we will use to
interpret our findings. In the superfluid regime ($U_b/J_b \ll 1$) the Bose gas supports
excitations in the form of Bogoliubov phonons, whose excitation spectrum is
given by~\cite{Oosten2001}
\begin{subequations}\label{eq:spectra}
\begin{align}
\hbar \omega_k &= \sqrt{ \varepsilon_k ( \varepsilon_k + 2U_b n_b)}, \label{eq:sfspectrum} \\
\varepsilon_k &= 2 J_b(1-\mathrm{cos}ka), \label{eq:tgspectrum}
\end{align}
\end{subequations}
where $n_b$ is the ratio of bosons to lattice sites. From this we find that
the density of states $( \hbar \partial \omega_k / \partial k)^{-1}$
diverges at the upper edge of the band with bandwidth $B = 4 J_b \sqrt{1 + U_b n_b / 2 J_b }$.
In the opposite limit $U_b/J_b \rightarrow \infty$ the bosons map to the
same number of non-interacting identical fermions with energy spectrum
$\varepsilon_k$. Hence in this regime the Bose gas supports particle-hole
excitations for $n_b < 1$ within the single band of bandwidth $B = 4J_b$.

For the commensurately filled case of $n_b = 1$ an increasing interaction
strength $U_b / J_b$ takes the gas through a continuous superfluid to Mott
insulator transition, which occurs at $U_b / J_b =
3.37$~\cite{Kuhner2000,Kollath2004} in the thermodynamic limit. A principal
signature of this transition is the appearance
of a gapped excitation spectrum suppressing the low-energy response. Deep in
the Mott insulator regime ($U_b/J_b \gg 1$) the gap in the excitation spectrum is $G \simeq
U_b$.
%%%%%%%%%%%%%%%%%%%%%%

\subsection{Detailed model and measuring procedure}
To summarize, the model for our system is given by the impurity
Hamiltonian $\hat{H}_a$, the Bose-Hubbard model $\hat{H}_b$ and the
density-density interaction $\hat{H}_{ab}$, as shown schematically in
\fir{fig:model}. More precisely, we consider an $M$-site system with
box boundary conditions or a suitably flat bottomed potential (see \cite{Mayrath2005} for an example of a box trap realized in an experiment). As in
\cite{Bruderer2010} we expect our system to share its bulk behavior with the
experimentally important case of an additional harmonic trap, provided that
the trap is sufficiently shallow. A realization of the model may also
draw on recent experimental successes in trapping atoms in species-specific
optical lattices~\cite{Catani2009,Lamporesi2010,Catani2011}.

The system is identical to that considered in~\cite{Bruderer2010} which
focussed on a superfluid environment ($U_b/J_b \ll 1$), whereby the impurity
and Bose gas together mimicked the electron-phonon interaction. However, the
control and flexibility of an optical lattice setup allows for the
creation of a strongly-interacting Bose gas ($U_b/J_b \gg 1$), the regime we
explore in this article.

To measure the impurity current we propose the following procedure: The $N$ bosons are cooled to their ground state with the impurities fixed in highly localized states at their initial locations (for example, using a tight optical trap). The system reaches equilibrium and the presence of the impurities inevitably leads to
density variations in the initial state of the bosons as compared to the ground state of the Bose-Hubbard Hamiltonian. At time $t = 0$
the impurities are released, their lattice is tilted and the system is left
to evolve. Later, a time of flight measurement~\cite{Bloch2008} or in situ
image~\cite{Bakr2009,Sherson2010} is taken~\footnote{Over multiple
realizations this allows the calculation of the elements of the reduced
single-particle density matrix in the Wannier basis $\langle \cre{a}_i
\an{a}_j \rangle$ or its diagonal, the density distribution.}. Note that
alternative preparations of the initial state we considered, e.g.,
preparing the bosons in equilibrium without the impurities and then adding
the impurities at $t = 0$, produced the same qualitative results despite
this being a more energetic initial state.

For our numerical simulations we consider the complete
two-species system governed by the Hamiltonian $\hat{H}_a + \hat{H}_b + \hat{H}_{ab}$. The simulations of the
full many-body quantum dynamics at zero temperature are performed using the
TEBD algorithm. This is a matrix product based method extending the powerful density matrix renormalization group to real time evolution. We present an overview of the method in \apr{sec:TEBD} along
with the algorithm parameters used.

We consider a single impurity initially localized in a Wannier state at
the center of the lattice. We evolve the system up to $t_{\mathrm{sim}}
\approx M \hbar / 4 J_b$, the time it approximately takes the disturbances
in the bath caused by the impurity to reach the boundaries of the system,
thus minimizing finite size effects. This time limit is determined by
the maximum group velocity $v_g$ of the bosons in the band which can be
derived in both the superfluid $U_b/J_b \rightarrow 0$ and
strongly-interacting $U_b/J_b \rightarrow \infty$ limits by differentiating
$\hbar\omega_k$ and $\varepsilon_k$ [Eqns. (\ref{eq:spectra})], respectively, with respect to $k$.
The time taken for disturbances in the Bose gas to travel a distance $Ma/2$
to the edge of the system is then $Ma/2v_g$ from which the previous
expression for $t_{\mathrm{sim}}$ follows. We choose to hold the interaction
between the impurity and the bosons constant at $U_{ab}/J_b = 0.5$ and the
impurity hopping at $J_a/J_b = 0.6$. These values are chosen to reflect the behavior we observed over a wide range. Notably they are outside the perturbative regime used in \secr{sec:setup} yet still lead to the predicted effects.
%%%%%%%%%%%%%%%%%%%%%%

\section{Simulations of impurity propagation}
\label{sec:current} 
%%%%%%%%%%%%%%%%%%%%%%
\begin{figure}[tbh]
\includegraphics[width=8 cm]{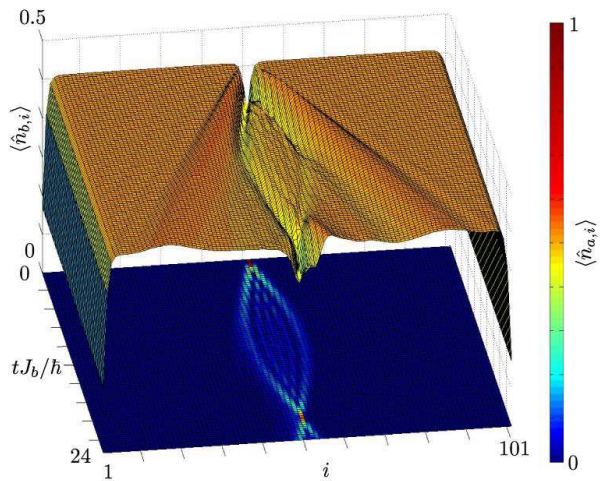}
\caption{\label{fig:numerical_example}(Color online) {\em The transport of
an impurity through a bosonic bath}. An example of the evolution of the
impurity and boson densities for the case $\Delta/J_b = 0.3$, $U_b/J_b = 2$,
$M=101$ and $N=50$. The color map in the $xy$-plane shows the expected
impurity density at each lattice site $\langle \hat{n}_{a,i} \rangle$, and
similarly for the surface plot and boson density $\langle \hat{n}_{b,i}
\rangle$.}
\end{figure}
%%%%%%%%%%%%%%%%%%%%%%

As a demonstration of the system and numerical method we
show the evolution of both the impurity and bosons in
\fir{fig:numerical_example} for an intermediate boson-boson interaction
strength $U_b/J_b = 2$. The impurity undergoes Bloch oscillations as it
would if the bosons were not present --- see \fir{fig:blochesakitsu}(a) --- however,
the non-zero impurity-boson interactions cause the impurity to dissipate
energy into the Bose gas and drift down the lattice. The same interactions
also create density fluctuations in the Bose gas, which spread out and reach
the system boundary near the end of the simulation.

Examining the system for a range of tilts $\Delta$ allows us to find the
dependence of impurity displacement $\langle \hat{x}_a \rangle = \sum_i a
(i_0 - i) \langle \hat{n}_{b,i} \rangle$ on $\Delta$, where $i_0$ is the
site at which the impurity was initially localized. Using this we can
calculate the average impurity velocity $\langle v \rangle = \langle \hat{x}_a
\rangle / t_{\mathrm{sim}}$. From detailed examination of our simulations we
find that the displacement does not increase just linearly in time but
oscillates slightly at the frequency of the Bloch oscillations (as would be
observed in an experiment). For all but the lowest tilts the durations of
evolutions considered is sufficient that by $t_{\mathrm{sim}}$ the
displacement has averaged over enough Bloch oscillations that $\langle v
\rangle$ gives a good representation of the long-time average drift.
%%%%%%%%%%%%%%%%%%%%%%

\subsection{Superfluid and Tonks-Girardeau regimes}
\label{sec:esakitsu}
%%%%%%%%%%%%%%%%%%%%%%
\begin{figure}[tbh]
\includegraphics[width=8 cm]{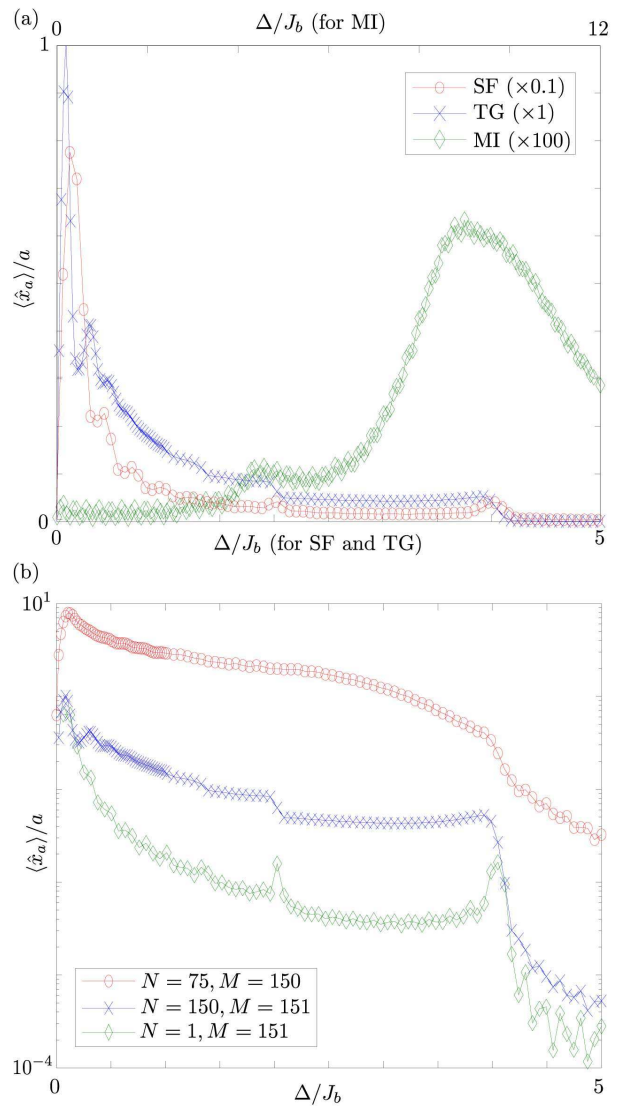}
\caption{\label{fig:currvstilt}(Color online) {\em Determining the bandwidth and gap from the dependence of displacement on tilt}. (a) The dependence of the displacement of the impurity $\langle
\hat{x}_a \rangle$ at $t_{\mathrm{sim}}$ as a function of tilt $\Delta$ for three regimes of the
Bose gas: the superfluid regime $U_b/J_b = 0.1$ with $N = 50$, $M = 101$
(SF - displacement scaled down by a factor of $0.1$); the Tonks-Girardeau
limit $U_b/J_b \rightarrow \infty$ with $N = 150, M = 151$ (TG); and the Mott insulating regime
$U_b/J_b = 10$ with $N = M = 101$ (MI -
displacement scaled up by a factor of $100$). (b) The same quantity is plotted  for several bosonic systems in the
Tonks-Girardeau limit, demonstrating that the qualitative shape and the
positions of the bandwidth suppressions are independent of the filling (so long as it is incommensurate) of
the Bose gas.}
\end{figure}
%%%%%%%%%%%%%%%%%%%%%%

Our starting point is the propagation of the impurity through a Bose gas in
the superfluid regime with incommensurate filling. A typical dependence of
the displacement of the impurity $\langle \hat{x}_a \rangle$ on the lattice
tilt $\Delta$ is given by the red circles in \fir{fig:currvstilt}(a).
This dependence has the characteristic features of the Esaki-Tsu result, shown in \fir{fig:blochesakitsu}(b), namely an ohmic region followed by negative
differential conductance.

However, a phenomenon that is specific to the case where the bosons feel a
lattice is the occurrence of resonances in the vicinity of the lattice tilts
$\Delta = 4J_b/\ell$ for integer $\ell \geq 1$. As shown in
\cite{Bruderer2010} we expect resonances at tilts $\Delta = (4 J_b /\ell )
\sqrt{1 + U_b n_b  / 2 J_b}$ corresponding to a divergence in the density of
phonon states at the upper band edge. Moreover, the suppression of the
propagation of the impurity for $\Delta\gtrsim 4J_b / \ell$ reflects the sudden
inability of the impurity to move $\ell$ sites down the lattice by
dissipating the energy $\ell\Delta$ into the Bose gas through a single
phonon process. Hence the superfluid Bose gas is a prime example of how the
value of the bandwidth of its excitation spectrum $B$ manifests itself in the
dependence of the current of the impurity. Note that the small fluctuations in the dependence of drift on tilt at small tilts in \fir{fig:currvstilt}(a) are a result of the finite time of our simulations and would be smoothed out by considering longer times. The same would be true in an experimental realization. 

One expects to see the bandwidth dependent features even with strong boson-boson
interactions; the reasoning used to predict current drop offs after $\Delta = B
/\ell$ in \secr{sec:setup} assumes the bosonic system is susceptible to low
energy excitations, which is the case for incommensurately
filled systems of any interaction strength~\cite{Oosten2001}. Indeed, we have simulated $M=25$ site systems over a range of boson-boson interaction strengths $U_b / J_b$ (not shown here) and found no qualitative change in the behavior of the motion. However, as shown in
\fir{fig:currvsint}, the average displacement of the
impurity decreases with increasing interaction strength. 

In the extreme case of infinitely strong interactions
$U_b/J_b\rightarrow\infty$, i.e., in the Tonks-Girardeau limit, the
situation simplifies considerably. The bosons map to identical
non-interacting fermions with bandwidth $B = 4 J_b$, and so the creation of
particle-holes at the Fermi surface provides low-energy excitations.
Accordingly, we expect a large impurity displacement at low lattice tilts
and sharp drops in propagation for tilts above integer divisions of the
bandwidth $4J_b/ \ell$, in agreement with \fir{fig:currvstilt}(a).

We find the same behavior in the Tonks-Girardeau limit for many different
bosonic filling factors, as shown in \fir{fig:currvstilt}(b). The
half-filled bosonic system gives rise to a peak impurity current that is an order of
magnitude above that for the cases where the bosonic lattice is filled by a
single or $M-1$ bosons. This can be attributed to there being a greater number of
excitations available in the half-filled bosonic system; using the fermionic
mapping we find there are of approximately $M^2/4$ single quasi-particle
excitations available in the half-filled case as compared to $M-1$ when
there is only one boson or $M-1$ present. This reasoning suggests that the
peak displacement may be used as a measure of the filling of the bosonic
system.
%%%%%%%%%%%%%%%%%%%%%%
\begin{figure}[tb]
\includegraphics[width=8 cm]{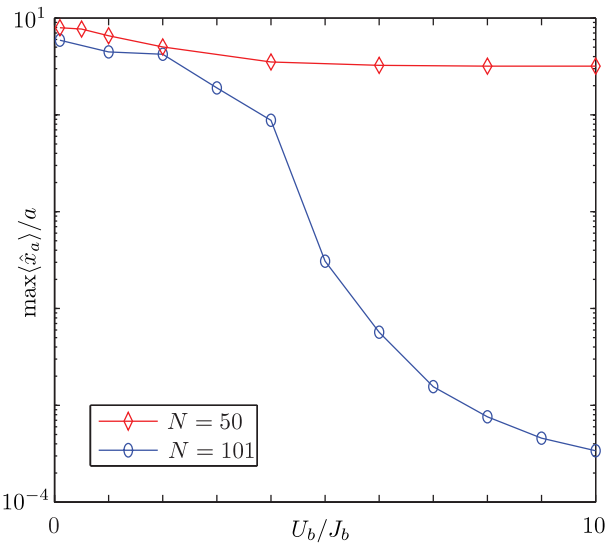}
\caption{\label{fig:currvsint}(Color online) {\em The suppression of average drift in the strongly-interacting regimes}. Here we show the peak displacement (maximized over all tilts) as a function of
interaction strength $U_b/J_b$ for commensurate and incommensurate fillings
of a $101$-site bosonic lattice.}
\end{figure}
%%%%%%%%%%%%%%%%%%%%%%

\subsection{Superfluid to Mott insulator transition}
\label{sec:sftomi} We now turn to the case of a commensurately filled Bose gas and the continuous superfluid to
Mott insulator transition. Since motion of the impurity though the
superfluid at low lattice tilts is a result of the ability of the bosons to
accept low-energy excitations we expect propagation to be highly suppressed
in the Mott insulator phase which has a gap in the excitation spectrum of $G
\simeq U_b$.

The effect of this signature on the propagation can be seen in
\fir{fig:currvstilt}(a), where the displacement has a peak at $\Delta
\approx U_b$. There is also a second peak at $\Delta \approx U_b /2$
resulting from second-order processes involving two lattice-site jumps,
similar to what is seen using Bragg spectroscopy in~\cite{Kollath2006} and
tilting the bosonic lattice in~\cite{Greiner2002}. As in those cases the
non-zero $J_b$ results in a broadening of the peaks. This behavior confirms
the predictions in \secr{sec:setup} for the transport of an impurity in a
gapped environment. This drastic change from the behavior predicted by Esaki and Tsu also tells us that the
relaxation-time approximation is not at all valid for the interaction of an
impurity with a Mott insulator.

For small tilts the motion of the impurity caused by
low-frequency excitations is completely suppressed; at large enough $U_b /
J_b$ the ground state is well approximated by a unit-filled Fock state. In
such a regime the impurity Bloch oscillates as it has no means of
dissipating energy into the Bose gas, hence there is no net current across
the lattice. As a result we get a startlingly clear signature of the
superfluid to Mott insulator transition in our dynamics in the form of a
quench of more than $4$ orders of magnitude in the propagation of the
impurity. This is shown in \fir{fig:currvsint}.

As well as the ability to probe the superfluid to Mott insulator transition
using the impurity, this quench provides a non-destructive way of
establishing whether the bosonic system is commensurately filled or not. If
our system is strongly interacting $U_b / J_b \gtrsim 4$ there are several
orders of magnitude separating the currents for an impurity in the
commensurate and incommensurate cases, shown in
\fir{fig:currvsint}. Many implementations of quantum information
processing involve creating a commensurately filled system. Using an
impurity as a probe is one way in which this commensurability could be
checked non-destructively.
%%%%%%%%%%%%%%%%%%%%%%
\begin{figure}[tbh]
\includegraphics[width=8 cm]{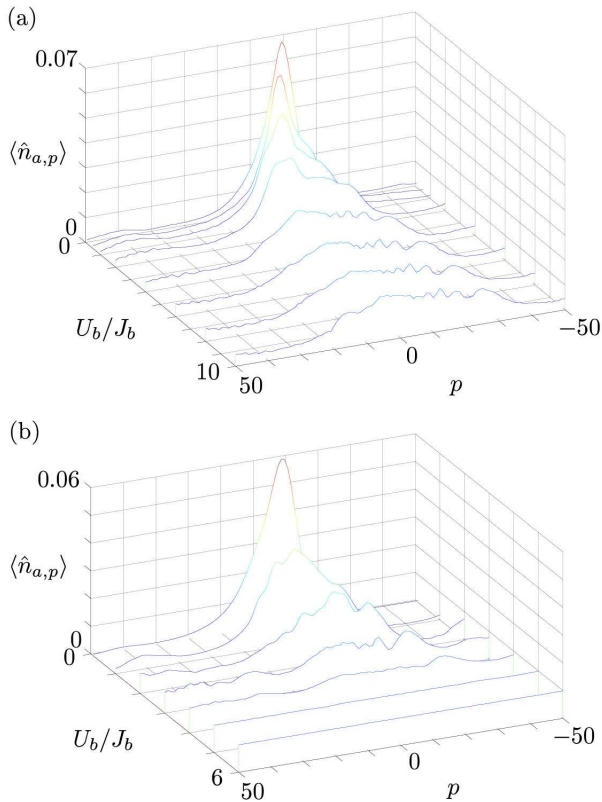}
\caption{\label{fig:momdist}(Color online) {\em Quasi-momentum distribution of the impurity after traveling through the Bose gas}. Here $\Delta / J_b
=
0.063$, $\hat{n}_{a,p}$ is the number operator for an impurity quasi-momentum of $2 \pi p / aM$ and $ M = 101$. (a) The momentum distribution of the
impurity at $t_{\mathrm{sim}}$ for a range of interaction strengths $U_b/J_b$ and incommensurate filling $N=50$. (b) The same as (a) but for
commensurate filling $N=101$.}
\end{figure}
%%%%%%%%%%%%%%%%%%%%%%

\subsection{Quasi-momentum distribution of the impurity} \label{sec:momdist}
Alternatively, one can gain information about the impurity by looking at
its quasi-momentum distribution, which is directly accessible in
time-of-flight experiments. With the impurity prepared in a Wannier state
each quasi-momentum state is equally occupied. However, we expect interactions with the Bose gas to alter this momentum distribution, resulting in the drift seen in \secr{sec:current}. The current $\langle v \rangle$ can be
calculated from the distribution in the semi-classical picture using \eqr{eq:semiclassicalcurrent}.

As can be seen in Figs. \ref{fig:momdist}(a) and \ref{fig:momdist}(b) the result of
interactions with the Bose gas is the enhancement of the distribution near the centre of the Brillouin zone. Moreover, the interaction gives rise to an asymmetry with the distribution skewed towards negative quasi-momentum, necessary for the impurity to have a current.

Our simulations show that a well-defined
quasi-momentum emerges in the superfluid regime by the time $t_{\mathrm{sim}}$, supporting the accuracy of
the semi-classical approximation. However, for stronger bosonic interaction
strengths the scattering results in a wide distribution with a negative quasi-momentum shoulder. By $U_b / J_b$ of the order $10$ the distribution already approximates its $U_b / J_b \rightarrow \infty$ value. In the case of the Mott insulating regime in \fir{fig:momdist}(b) it is
unchanged from its initially flat distribution. Once again, the momentum distribution of the impurity provides a signature
of the transition (cross-over) between the superfluid and Mott insulator
(Tonks-Girardeau) regimes.

The difference in steady-state quasi-momentum distribution is related to the structure factor of the Bose gas, c.f.
Eqns. (\ref{eq:boltzmanneq}) and (\ref{eq:fgr4}), but a quantitive
comparison with our simulations is difficult since in general they do not reach the steady-state. Since this limitation is due to the finite time of our simulations it would also restrict an experimental realization.
%%%%%%%%%%%%%%%%%%%%%%
\begin{figure}[tbh]
\includegraphics[width=8 cm]{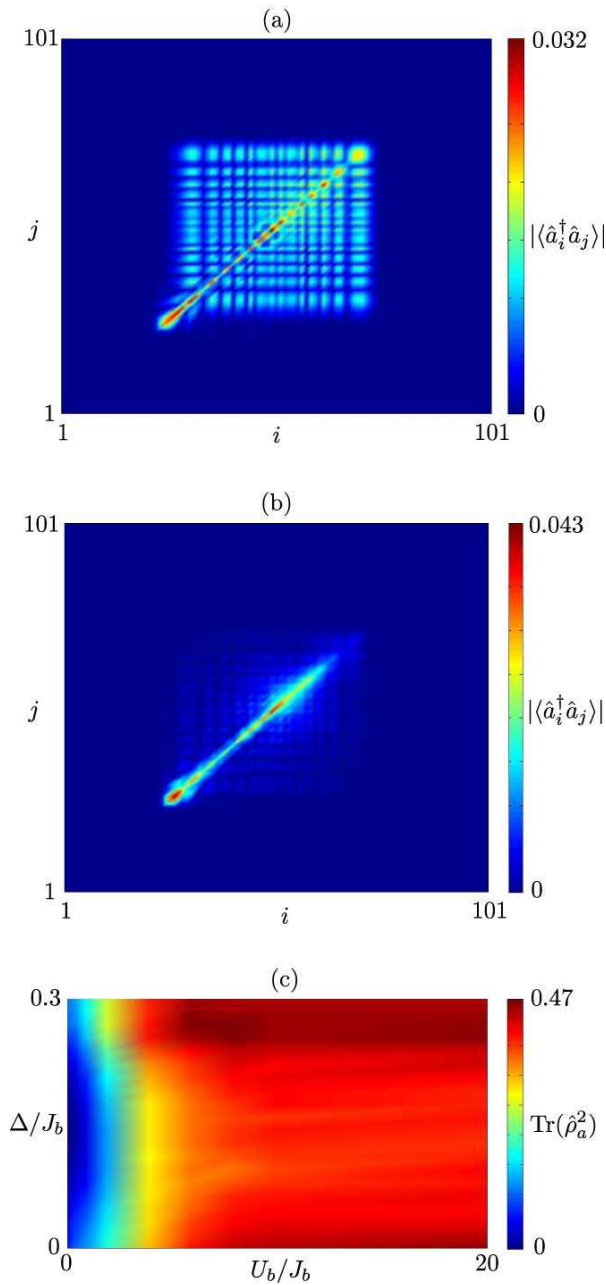}
\caption{\label{fig:denmat}(Color online) {\em Decoherence of an impurity traveling through an incommensurately filled system $N = 50, M = 101$}. (a)
Absolute
values of the elements of the reduced density matrix for the impurity $| \langle \cre{a}_i \an{a}_j \rangle |$ with $U_b/J_b = 2$ and $\Delta/J_b =
0.1$, calculated
at $t_{\mathrm{sim}}$. (b) The same as (a) but for $U_b/J_b = 0.1$. (c) Also at $t_{\mathrm{sim}}$ we have plotted the purity of the density matrix
$\Tr ( \hat{\rho}^2_a )$ as a function of tilt and interaction strength.}
\end{figure}
%%%%%%%%%%%%%%%%%%%%%%

\subsection{Reduced density matrix of the impurity} \label{sec:denmat} So far we have focused on the
evolution of the spatial and quasi-momentum distributions of the impurity, however more
information is contained in its reduced density matrix $\hat{\rho}_a$. First, evaluating the elements of the density matrix in the Wannier basis $\langle \cre{a}_i \an{a}_j \rangle$ may tell us more about the effect
of boson-impurity interactions on the impurity. Another quantity that can be calculated from the reduced density matrix is the purity $\Tr(\hat{\rho}^2_a)$, which takes the
maximal value $1$ if $\hat{\rho}_a$ describes a pure state and decreases for increasingly mixed
states. The initial state of the total system is a product of pure states for the impurity
and bosons, and since we consider coherent evolution it remains pure. However, through interactions of the impurity with the Bose gas the two subsystems are entangled resulting in a loss of purity of
the impurity. Our bosonic environment is sufficiently large that for most parameter values revivals are not observed and so we use purity to quantify the decoherence of the impurity.

To demonstrate the behavior of the density matrix $\hat{\rho}_a$ we have
plotted in Figs. \ref{fig:denmat}(a) and \ref{fig:denmat}(b) the absolute
values of its Wannier basis elements at $t_{\mathrm{sim}}$ for both intermediate and weak bosonic interaction strengths,
respectively. In Fig.~\ref{fig:denmat}(a) we observe a symmetric
coherent interference pattern in the off-diagonal elements as the impurity
undergoes Bloch oscillations. Additionally, density-density interactions with the
bosons result in a slight decay of the off-diagonal coherences and an asymmetric distribution along the diagonal with a higher occupancy of the lower
energy lattice sites. This decoherence is much more pronounced for a
superfluid environment, as shown in \fir{fig:denmat}(b), and the resulting
density matrix is nearly diagonal.

To make this relationship between the decoherence of the impurity and the
internal interaction strength of the environment clearer,
\fir{fig:denmat}(c) shows the purity of the reduced density matrix of the impurity as a function of
both tilt $\Delta$ and interaction strength $U_b / J_b$. As can be seen in
the figure, the coherence of the impurity is a clear signature of the
cross-over between the superfluid and Tonks-Girardeau regimes. Moreover, the values of $U_b / J_b$ at which the most significant changes occur is around the same values where the transition occurs in a commensurate Bose gas. We also calculated the entanglement entropy between the impurity and Bose gas (not shown here) which showed very similar behavior to purity. If the Bose gas is commensurately filled the transition between weak and strong interaction strengths makes an even clearer impact on the purity as deep in the Mott insulator phase the impurity does not become at all entangled with the Bose gas.

Finally, comparing Figs. \ref{fig:denmat}(c) and \ref{fig:numerical_example}(d) we
find that the regimes of high transport coincide with those of the smallest purity; the very same scattering processes that lead to the propagation of
the impurity along the lattice also cause its decoherence. Therefore the
propagation of the impurity provides a controllable example of noise-enhanced transport. Noise-enhanced transport is a topic of interest for those
searching for quantum effects at room temperature which explain the
remarkable efficiency of some biological systems~\cite{Plenio2008}.
%%%%%%%%%%%%%%%%%%%%%%

\section{Conclusions}
\label{sec:conclusion} We performed fully quantum many-body simulations of
large (100 or more sites) two-species systems using the TEBD algorithm,
simulating the one-dimensional motion of an impurity confined to a tilted
optical lattice through a Bose lattice gas. We discussed the Esaki-Tsu
dependence of the current of the impurity on tilt and provided evidence that this shape is qualitatively obeyed by an impurity traveling through an incommensurately filled bosonic system of any interaction strength including the Tonks-Girardeau limit. Contrasting this a very different dependence was found in the Mott insulating regime. 

We discussed how deviations to the Esaki-Tsu shape depend on properties of the Bose gas. Information about the gas can then be extracted without needing to measure the bosons themselves,
i.e., non-destructively. In particular we discussed how the impurity could
be used to probe the commensurateness of the system, and obtain the gap and bandwidth
of its excitation spectrum.

Following this we analyzed the transport of the impurity by considering its momentum distribution, showing that the semi-classical picture of an impurity of well-defined quasi-momentum emerges after a short time for bosons in the superfluid regime. We have also found a relationship between the decoherence and transport
of the impurity; our setup provides an experimentally accessible prototype of noise enhanced transport. 
%%%%%%%%%%%%%%%%%%%%%%

\appendix
%%%%%%%%%%%%%%%%%%%%%%

\section{Time-evolving block decimation}
\label{sec:TEBD} The TEBD algorithm efficiently simulates both the unitary
and imaginary time-evolution (according to $\eh^{-i\hat{H}t}$ and
$\eh^{-\hat{H}t}$ respectively) of quantum systems comprising a one-dimensional
lattice of subsystems, each with a finite number of configurations, where
$\hat{H}$ is composed of nearest neighbor terms~\cite{Vidal2003,Vidal2004}.
It does this by storing the state of a system as a matrix product state
(MPS) of dimension $\chi$. In general, for an exact representation of the
state vector the dimension $\chi$ and hence the resources required for the
simulation must grow exponentially with the size of the lattice. However for
ground and low-lying excited states of one-dimensional systems near-exact
accuracy can be obtained using a much smaller $\chi$ that does not grow
exponentially with system size~\cite{Verstraete2006,Eisert2010}. With this
being the case and using a Suzuki-Trotter decomposition of the operator
$\eh^{-i \hat{H} \delta t}$ (without the $i$ for imaginary
time-evolution)~\cite{Suzuki1990} the evolution may be simulated by the
application of a polynomial number of two-site gates, each followed by a
re-compression into MPS form, which can be done efficiently. For details, we
refer the reader to~\cite{Johnson2010,Daley2004,Vidal2003,Vidal2004}.

Our system, described in \secr{sec:setup}, is of this type provided we
restrict the occupancy of the bosonic lattice to a finite number. To realize
our scheme we first used TEBD to find the ground state of the bosonic part of
the system, with an on-site potential $U_{ab}$ added at each site
corresponding to an impurity. The initial state of the whole system was then
the tensor product of this bosonic state with the initial state of the
impurity, which consists of Fock states with occupancies 0 or 1. Next, TEBD
was used to simulate the evolution of this state under the full
Hamiltonian $\hat{H} = \hat{H}_{a} + \hat{H}_{b} + \hat{H}_{ab}$.
Expectation values such as bosonic and impurity densities were easily
extracted from the numerics due to the efficient contractibility of an MPS.

For our simulations we used a maximum boson lattice-site occupancy of $4$
and the largest MPS dimension we used was $\chi = 120$. Our timestep was
$\delta t = 5 \times 10^{-3} \hbar / J_b$. These parameters were sufficient for the time-evolution considered here; we found that increasing $\chi$ and decreasing $\delta t$ resulted in no significant changes to the observables computed.
%%%%%%%%%%%%%%%%%%%%%%

\acknowledgements THJ thanks Vlatko Vedral for interesting discussions and inviting him to CQT Singapore
where part of this work was completed. THJ and SRC thank John Goold for interesting discussions. MB thanks the Swiss National
Science Foundation for the support through the project PBSKP2/130366. SRC
and DJ thank the National Research Foundation and
the Ministry of Education of Singapore for support.
%%%%%%%%%%%%%%%%%%%%%%

\end{document}